\documentclass[10pt,journal,compsoc]{IEEEtran}
\usepackage{amsmath,amsfonts}
\usepackage{algorithmic}
\usepackage{algorithm}
\usepackage{array}
\usepackage[subtle]{savetrees}
\usepackage[caption=false,font=normalsize,labelfont=sf,textfont=sf]{subfig}
\usepackage{textcomp}
\usepackage{stfloats}
\usepackage{url}
\usepackage{verbatim}
\usepackage{graphicx}
\usepackage{cite}
\usepackage{xcolor}
\usepackage{hyperref}
\usepackage{booktabs}
\usepackage{fix-cm}

\hypersetup{
    colorlinks=true,
    citecolor=magenta,
    linkcolor=black,
    urlcolor=blue
}
\usepackage{tikz}
\begin{document}

\title{Engineering Trustworthy Machine-Learning Operations with Zero-Knowledge Proofs}

\author{Filippo~Scaramuzza,~\IEEEmembership{Member,~IEEE,}
        Giovanni~Quattrocchi,~\IEEEmembership{Member,~IEEE,}
        and~Damian~A.~Tamburri,~\IEEEmembership{Member,~IEEE}
\thanks{F. Scaramuzza is with Tilburg University, Tilburg, The Netherlands, and Jheronimus Academy of Data Science, 's-Hertogenbosch, The Netherlands (e-mail: f.scaramuzza@tilburguniversity.edu).}
\thanks{G. Quattrocchi is with Politecnico di Milano, Milan, Italy (e-mail: giovanni.quattrocchi@polimi.it).}
\thanks{D. A. Tamburri is with Università del Sannio, Benevento, Italy, Jheronimus Academy of Data Science, 's-Hertogenbosch, The Netherlands, and NXP Semiconductors, Eindhoven, The Netherlands (e-mail: datamburri@unisannio.it).}
}


\markboth{IEEE Transactions on Software Engineering to appear, 2026}%
{{Scaramuzza} \MakeLowercase{\textit{et al.}}: IEEE Transactions on Software Engineering}

\IEEEpubid{0000--0000/00\$00.00~\copyright~2021 IEEE}

\IEEEtitleabstractindextext{%
\begin{abstract}
As Artificial Intelligence (AI) systems, particularly those based on machine learning (ML), become integral to high-stakes applications, their probabilistic and opaque nature poses significant challenges to traditional verification and validation methods. These challenges are exacerbated in regulated sectors requiring tamper-proof, auditable evidence, as highlighted by apposite legal frameworks, e.g., the EU AI Act. Conversely, Zero-Knowledge Proofs (ZKPs) offer a cryptographic solution that enables provers to demonstrate, through verified computations, adherence to set requirements without revealing sensitive model details or data. Through a systematic survey of ZKP protocols, we identify five key properties (non-interactivity, transparent setup, standard representations, succinctness, and post-quantum security) critical for their application in AI validation and verification pipelines. Subsequently, we perform a follow-up systematic survey analyzing ZKP-enhanced ML applications across an adaptation of the Team Data Science Process (TDSP) model (Data \& Preprocessing, Training \& Offline Metrics, Inference, and Online Metrics), detailing verification objectives, ML models, and adopted protocols. Our findings indicate that current research on ZKP-Enhanced ML primarily focuses on inference verification, while the data preprocessing and training stages remain underexplored. Most notably, our analysis identifies a significant convergence within the research domain toward the development of a unified Zero-Knowledge Machine Learning Operations (ZKMLOps) framework. This emerging framework leverages ZKPs to provide robust cryptographic guarantees of correctness, integrity, and privacy, thereby promoting enhanced accountability, transparency, and compliance with Trustworthy AI principles.
\end{abstract}

}
 
\maketitle
\IEEEdisplaynontitleabstractindextext
\IEEEpeerreviewmaketitle

\section{Introduction}
\label{sec:introduction}
Artificial Intelligence (AI) software has become a critical component in numerous applications, ranging from autonomous driving~\cite{9478090} and healthcare diagnostics~\cite{9676233} to financial decision-making and public service automation~\cite{KUZIEMSKI2020101976}. The rapid advancement and adoption of AI technologies have brought profound benefits, but also significant challenges related to reliability, safety, and ethics. As AI systems increasingly influence high-stakes domains, ensuring their trustworthiness and robustness is essential~\cite{kaur2022trustworthy}. One of the key processes to establish trust is software verification and validation, which aims to demonstrate that a software system meets its declared properties and performs as expected under realistic operating conditions~\cite{wallace1989software}.

Traditionally, software verification and validation have relied on a combination of testing, static analysis, and documentation-based processes such as performance reports, external audits, and model cards~\cite{Myllyaho2021Systematic}. While these approaches have proven effective for conventional software, they face significant limitations when applied to AI systems, particularly those based on machine learning (ML). ML models are inherently probabilistic, data-dependent, and often opaque, complicating the assessment of correctness and compliance. Furthermore, the deployment of ML models as services (MLaaS)~\cite{Groot2021Availability} introduces additional challenges, as the model internals remain inaccessible to external validators. This black-box nature limits direct inspection and complicates verification of whether the declared model was actually used for inference, or whether reported performance metrics truthfully represent the deployed system’s behavior~\cite{Adler2016Auditing}. Consequently, traditional validation approaches struggle to provide objective, tamper-proof evidence, weakening accountability and trust, especially in regulated sectors where compliance mandates clear, auditable validation evidence, as emphasized by recent legislation such as the EU AI Act~\cite{EUAIAct2023}.

A promising approach to improve validation transparency and objectivity is the use of \emph{Zero-Knowledge Proofs} (ZKPs)~\cite{goldwasser2019knowledge}. ZKPs are cryptographic protocols that allow one party (the prover) to demonstrate to another party (the verifier) that a computation was carried out correctly, without requiring the verifier to rerun the computation or access sensitive internal details. Originally developed for the broader field of verifiable computing, ZKPs have increasingly been applied to ML, where, for example, they can offer a mechanism to prove that an inference step was executed correctly using a declared model, without revealing the model’s internal parameters or the input data itself~\cite{xing2023zero}.

This work focuses on evaluating the feasibility of applying ZKPs to the broader challenge of \emph{Trustworthy AI Software verification and validation in the MLOps lifecycle}. 

By embedding ZKPs into AI software workflows, it becomes possible to generate tamper-proof, cryptographically verifiable evidence that computations adhere to declared specifications and requirements, without revealing sensitive details such as proprietary model weights or training data. This approach enables external auditors, customers, or regulators to independently verify AI software operations while respecting intellectual property concerns.
In summary, the key contributions of this work are: (a) a systematic survey of ZKP protocols, highlighting five key properties (non-interactivity, transparent setup, standard representations, succinctness, and post-quantum security) that make them suitable for integration into AI system verification and validation pipelines; (b) a structured analysis of ZKP-enhanced ML applications, organized according to the stages of the TDSP model~\cite{amershi2019software}, and for each application, the specific verification objective, the ML model used, and the ZKP protocol adopted are detailed; (c) An exploration of the emerging convergence between ZKP and ML technologies toward a unified \emph{Zero-Knowledge Machine Learning Operations} (ZKMLOps) verification framework for Trustworthy AI, identifying research trends and future works.

The remainder of this paper is organized as follows. Section~\ref{sec:background} provides background on Trustworthy AI, AI software verification and validation, and Zero-Knowledge Proofs. Section~\ref{sec:methodology} outlines the research methodology. Sections~\ref{sec:zkp_protocols} presents a systematic literature review on ZKP protocols, identifying 5 key properties that make them suitable for integration into AI system verification and validation pipelines. Section~\ref{sec:zkml} presents a systematic literature review on ZKP-Enhanced ML applications, showing the convergence of the research domain toward a unified \emph{Zero-Knowledge Machine Learning Operations} (ZKMLOps) verification framework for Trustworthy AI. Section~\ref{sec:future_work} outlines potential research directions and opportunities for extending the contributions of this work.
Section~\ref{sec:conclusions} concludes the work, highlighting the key findings of the research.

\section{Background}
\label{sec:background}
This section lays the foundational groundwork, first by outlining the principles of Trustworthy AI, then by detailing the specific challenges in AI Software Verification and Validation, and finally by introducing Zero-Knowledge Proofs as the foundational cryptographic technique for this work.

\subsection{Trustworthy AI}
\emph{Trustworthy AI} has emerged as a critical area of focus as AI systems increasingly impact society, business, and everyday life. Ensuring that these systems are reliable, ethical, and safe is essential for promoting public trust and for enabling the responsible deployment of AI technologies at scale.

The concept of Trustworthy AI is rooted in five foundational ethical principles: beneficence, non-maleficence, autonomy, justice, and explicability~\cite{thiebesTrustworthyArtificialIntelligence2021}. 
There is a set of well-established technical and ethical dimensions of trustworthy AI~\cite{kaur2022trustworthy,liuTrustworthyAIComputational2022b}: (i) \emph{Safety \& Robustness}, i.e. ensuring systems perform reliably under various conditions, (ii) \emph{Fairness \& Non-discrimination}, i.e. preventing bias and ensuring equitable outcomes, (iii) \emph{Explainability \& Transparency}, i.e. making AI decisions understandable and traceable, (iv) \emph{Privacy \& Data Governance} protecting user data and ensuring responsible data use, (v) \emph{Accountability \& Auditability}, i.e. assigning responsibility and enabling oversight, (vi) \emph{Societal \& Environmental Well-being}, i.e. considering broader impacts on society and the environment.

A systematic approach to trustworthy AI spans the entire AI lifecycle, from data acquisition and model development to deployment and monitoring, and includes the following key components~\cite{liTrustworthyAIPrinciples2023c,diaz-rodriguez_connecting_2023}:
(i) \emph{Risk Analysis}, i.e., identifying and mitigating potential ethical, technical, and societal risks.
(ii) \emph{Validation}, i.e,. ensuring the AI system meets performance goals and stakeholder expectations in its intended context, 
(iii) \emph{Verification}, i.e., confirming that the system adheres to design specifications and functions as intended,
(iv) \emph{Continuous Governance}, i.e,. maintaining oversight to ensure long-term accountability, compliance, and adaptability.

\subsection{AI Software Verification and Validation}
Software validation is a well-established process in traditional software engineering, ensuring that software fulfills its declared requirements and performs as intended~\cite{wallace1989software}.
When applied to AI software, validation becomes significantly more challenging. Traditional validation techniques assume deterministic behavior, where outputs are traceable to explicitly written source code. Modern AI systems, especially those based on ML, exhibit probabilistic behavior that depends heavily on training data, model architecture, and optimization processes. This makes it harder to directly link observed outputs to the intended requirements~\cite{Myllyaho2021Systematic}. Further complicating the process, many AI models are proprietary and deployed as services, meaning external validators, regulators, or customers cannot access the internal details of the model. This black-box nature forces external parties to rely on documentation or self-reported performance metrics, limiting the objectivity and reproducibility of the validation process. Moreover, current approaches such as model cards or empirical performance reports provide useful context, but they are fundamentally self-declared and do not inherently provide verifiable evidence~\cite{Myllyaho2021Systematic}. In turn, external validation mechanisms, such as audits or independent re-testing, also face practical limits when applied to AI systems. Audits rely on documentation provided by the developer, creating risks of selective reporting. Independent re-testing, while more objective, may be infeasible for large or proprietary models where data and models cannot be freely shared~\cite{casper2024black}.

\subsection{Zero-Knowledge Proofs}
ZKPs provide a formal mechanism through which a \emph{prover} can convince a \emph{verifier} that a given statement is true, without revealing any information beyond the truth of the statement itself~\cite{goldreich2001foundations}. 

To introduce the idea, consider a traditional software application used to determine eligibility for a benefit based on income. The rule might be: ``grant the benefit if the citizen's income is less than \$30,000.'' With a ZKP, the citizen (prover) can convince an organization (verifier) that their income satisfies this condition, without revealing the actual income. 

At the core of modern ZKP systems is the transformation of any arbitrary computations into \textit{arithmetic circuits} defined over finite fields \cite{goldwasser2019knowledge}. Any computable function can be rewritten as a sequence of additions and multiplications over a finite field $\mathbb{F}_p$, where $p$ is a large prime. The prover’s task is to demonstrate knowledge of a valid assignment to all the variables in the circuit, ensuring that all constraints hold. Formally, the prover proves the existence of a secret witness \(w\) that satisfies:
\[
C(x, w) = y
\]
where \(C\) denotes the arithmetic circuit, \(x\) represents public inputs, \(w\) is the private witness, and \(y\) is the public output of the computation. If we consider the previous example:
\begin{itemize}
    \item The public input~$x$ encodes the eligibility threshold (\$30,000).
    \item The witness~$w$ represents the citizen's confidential income.
    \item The public output~$y$ is the Boolean result (e.g., true if the condition holds).
\end{itemize}
The ZKP convinces the verifier that there exists a secret~$w$ such that the circuit $C$ satisfies $C(x, w) = y = \texttt{true}$, without revealing~$w$.

ZKPs were first studied in the setting of \emph{interactive proofs}~\cite{goldwasser2019knowledge}, where the prover and verifier engage in a sequence of challenge-response rounds. These protocols guarantee that a cheating prover cannot convince an honest verifier of a false statement, except with negligible probability. A significant step towards removing interaction was the \emph{Fiat-Shamir heuristic}~\cite{fiat1986prove}. This technique transforms certain interactive protocols into non-interactive variants by replacing the verifier’s random challenges with the output of a cryptographic hash function applied to the transcript. While widely used and practical, this transformation's security is typically proven in the idealized Random Oracle Model~\cite{bellare1993random}. Blum et al.~\cite{blum2019non} later gave a precise mathematical definition of \emph{Non-Interactive Zero-Knowledge Proofs} (NIZKs) and showed how to build them with provable security guarantees in the standard cryptographic model, typically using a shared reference string that all parties can access. Both approaches result in a self-contained proof that can be verified without further interaction.

To enable efficient proof generation and verification, many systems encode the execution trace of the computation into a polynomial~$P(x)$ over~$\mathbb{F}_p$:
\[
P(x) = \sum_{i=0}^{n} c_i x^i
\]
The prover commits to this polynomial using a \emph{polynomial commitment scheme}~\cite{kate2010constant}, which ensures both \emph{binding} (the committed polynomial cannot be altered later) and optionally \emph{hiding} (its content remains secret). The verifier can then check whether the polynomial satisfies the required properties by querying a few evaluations at selected points. This drastically reduces the size of the proof and the cost of verification, achieving the property of \emph{succinctness}.

A key challenge in applying ZKPs to domains such as ML is handling \emph{non-linear functions}, which are not naturally supported in arithmetic circuits. Neural networks, for example, often include non-linear activation functions like the Rectified Linear Unit ($ReLU(x) = \max(0, x)$)~\cite{arora2018understandingdeepneuralnetworks}.
To represent such operations in ZKP-friendly form, systems typically use \emph{lookup arguments}~\cite{kang2022scalingtrustlessdnninference}. In a lookup argument, the prover shows that each non-linear operation maps an input to an output according to a precomputed table~$T$:
\[
\exists (x, y) \in T \quad \text{such that} \quad y = f(x)
\]
This allows incorporating non-polynomial logic into ZKPs while preserving succinctness and zero-knowledge. The table~$T$ encodes valid input-output pairs for the non-linear function, and the verifier only checks that the prover’s values appear in the table.



\section{Related Work}
\label{sec:related_work}
To demonstrate the significance of our contribution, we conducted a comprehensive review of pertinent literature by examining leading conferences and journals, complemented by a snowballing methodology. We aimed to identify works that survey the applicability of ZKP protocols to ML, particularly those that delineate critical factors and properties, as well as studies exploring the integration of ZKP within ML applications. The review revealed several surveys, each addressing specific facets of ZKP in ML; however, none provided a holistic perspective on the integration of ZKP across the MLOps pipeline within the broader context of Trustworthy AI verification and validation.

Lavin et al.~\cite{lavin_survey_2024} present a comprehensive survey aimed at both researchers and practitioners, covering a wide spectrum of real-world applications and use cases of ZKPs. Within the domain of ML, the survey contextualizes recent advances---including those discussed in Section~\ref{sec:zkml} of this work---highlighting the current state of the art. While the contribution is substantial, it does not explicitly address the MLOps lifecycle nor provide an in-depth discussion of protocol-level considerations, ML model addressed, or verification processes essential to operationalizing ZKPs in ML pipelines. 

Peng et al.~\cite{peng_survey_2025} deliver a survey of Zero-Knowledge Machine Learning (ZKML) research, covering works from June 2017 to December 2024, which they categorize into verifiable training, inference, and testing, complemented by discussions on implementation challenges and commercial applications. Their work offers a valuable chronological and stage-based overview of the ZKML field. While comprehensive in its temporal scope and categorization by verification stage, the survey does not extend its analysis to a detailed mapping of ZKP-enhanced ML applications across a full MLOps lifecycle process. Furthermore, their review does not place a central focus on a systematic, criteria-driven assessment of ZKP protocol characteristics for AI system verification, nor on the explicit conceptualization of a unified MLOps framework designed to integrate ZKPs for advancing Trustworthy AI.
We found only one paper, by Balan et al.~\cite{balan2025framework}, that proposes a framework for verifiability across the whole AI pipeline. They identify key parts and link existing cryptographic tools to different stages, from data sourcing to unlearning, aiming to allow verification of AI-generated assets. While their goal of a complete view is valuable, the pipeline stages they describe (such as ``verification of raw dataset'' and ``extraction and analysis'') are presented generally and do not seem to follow a formal MLOps model. The authors also state that, as yet, ``there are no implementations of this fully verifiable pipeline,'' which shows such end-to-end solutions are still largely conceptual. Therefore, their work does not offer a systematic survey of existing ZKP-enhanced ML applications organized by a standard MLOps lifecycle, nor does it deeply analyze ZKP protocol suitability for various ML tasks using specific criteria—areas central to our contributions.

In summary, while the reviewed literature provides valuable insights into ZKP applications for ML, general ZKP surveys or conceptual frameworks for engineering AI verifiability with ZKP approaches are missing, which motivates our work in proposing a framework to provide a holistic approach to Trustworthy Machine Learning Operations with ZKPs. 
\section{Methodology}
\label{sec:methodology}

This work adopts a mixed methodology that combines two systematic literature reviews following the methodology described by Kitchenham et al.~\cite{kitchenham2009systematic} with a systematic analysis of ZKP protocols and their applications in ML. The first review identifies and characterizes relevant ZKP protocols, examining their mathematical foundations, performance properties, and implementation maturity. The goal is to identify common patterns and challenges and define a set of essential properties that a ZKP protocol should possess to be effectively applied in an ML context. The second review analyzes the emerging field of ZKP-Enhanced ML, exploring how ZKPs have been applied to validate and secure ML processes. We further classify each relevant contribution based on the \emph{Team Data Science Process} (TDSP) model~\cite{amershi2019software} to show the convergence of this research domain towards a unified MLOps pipeline verification framework.

Furthermore, to encourage replication, we provide a full replication package\footnote{\url{https://tinyurl.com/yc5snret}} available online.

\subsection{Literature Search Process for ZKP Protocols}

The first systematic literature review focused on identifying and characterizing the main ZKP protocols that could potentially be applied to inference validation in ML systems. Since this initial review was intended to capture the landscape of general-purpose ZKP protocols, its scope was not restricted to ML-specific applications, allowing for a broader understanding of available proof systems, their theoretical properties, and their practical characteristics.

\subsubsection{Research Query}
The query applied for this search was:
\begin{quote}
\texttt{("zero knowledge" OR "verifiable comput*") AND (proof OR argument) AND (interactive OR "non-interactive")}
\end{quote}
This query was designed to retrieve works that focus on both interactive and non-interactive proof systems, including both classical ZKPs and broader verifiable computing techniques. The search was performed in the ACM Digital Library\footnote{\url{https://dl.acm.org}}, IEEE Xplore\footnote{\url{https://ieeexplore.ieee.org}}, and Cryptology ePrint Archive\footnote{\url{https://eprint.iacr.org}}, as these libraries cover the main venues where ZKP research has been published.

\subsubsection{Screening and Filtering Process}
The search yielded a total of 1,427 papers across all three libraries. To refine this set, a comprehensive filtering process was applied, consisting of three main phases: title screening, abstract screening, and full-text assessment. In the title screening phase, papers were evaluated based on their titles, and those clearly indicating topics unrelated to the core focus of ZKP contributions---such as works exclusively centered on blockchain applications, finance, or other domains with no relevance to general ZKP advancements---were excluded. During the abstract screening phase, papers were further assessed to eliminate those that, despite referencing ZKPs, did not offer direct contributions to the design, analysis, or benchmarking of ZKP protocols. Additionally, duplicates across the libraries were identified and removed to ensure a unique set of studies. In the final phase, full-text assessment was conducted, where each remaining paper was thoroughly reviewed to confirm that it provided a meaningful discussion of ZKP protocols themselves, rather than merely applying pre-existing protocols to external use cases without novel insight. Papers failing to meet this criterion were discarded, and any remaining redundancies were addressed. After completing this rigorous process, a final set of 30 papers was obtained.

\subsubsection{Quality Indices}
To systematically assess the quality of these 30 papers, we defined a set of quality indices, inspired by established methodologies in literature reviews~\cite{7929422}. These indices evaluate key aspects of each study, assigning scores from 0 to 2 based on specific criteria, like problem definition, problem context, research design, results, insights derived, and limitations. Each surviving paper was thoroughly read and scored according to these metrics, which include the clarity of problem definition, the depth of contextual description, the explicitness of research design, the specificity of contributions, the insightfulness of derived lessons, and the acknowledgment of limitations. This scoring mechanism enabled us to prioritize papers that not only meet the thematic relevance criteria but also exhibit robustness and transparency in their scientific approach. The resulting quality scores provide a foundation for identifying the most significant works that shape our understanding of ZKP protocols and their theoretical advancements.

\subsection{Systematic Literature Review on ZKP-Enhanced ML}

The second component of the methodological process consisted of a systematic literature review SLR focused specifically on the intersection of ZKPs and ML. This review aimed to identify existing approaches where ZKPs were applied to ML processes. The objective was to understand how the current research landscape addresses the need for externally verifiable, privacy-preserving validation of ML computations.

\subsubsection{Research Query}

The following search query was developed to capture works focusing explicitly on the use of ZKPs for verifying or validating ML processes:
\begin{quote}
\texttt{("zero knowledge proof" OR "verifiable comput*") AND ("ML" OR "neural network" OR "deep learning")}
\end{quote}

This query was executed across two major digital libraries, IEEE Xplore and ACM Digital Library. The Cryptology ePrint Archive was excluded from this review as a pilot study showed a lack of directly relevant work focusing on ML inference.

\subsubsection{Screening and Filtering Process}

The initial query returned a total of 1,134 papers across the two libraries. These papers were filtered in two stages, applying progressively stricter criteria to ensure relevance to the topic of ZKP-enhanced ML validation. In the first stage, papers were excluded if they focused only on privacy-preserving ML techniques unrelated to ZKPs, or if they discussed general ML security (such as adversarial attacks or robustness) without addressing verification tasks.
The remaining papers underwent the second stage, which involved a full-text review, with papers excluded if they:
(i) Used ZKPs only as a theoretical reference without concrete implementation or application to ML workflows;
(ii) Incorporated ZKPs in ways that did not contribute to verifiability or correctness validation, such as merely enhancing privacy without any verification objective;
(iii) Applied existing ZKP protocols without modification or novel insight, offering limited contribution to the understanding or evolution of ZKP-Enhanced ML.

This process left a final set of 42 papers for inclusion in the literature review.

\subsubsection{Cross-Referencing and Snowballing}

To maximize coverage, an additional round of cross-referencing was conducted using the citations and bibliographies of the 42 selected papers. This step identified 15 additional works of relevance, bringing the final corpus to 57 papers.

\subsubsection{Comparative Analysis}

\begin{figure*}[th]
    \centering
    \includegraphics[width=\linewidth]{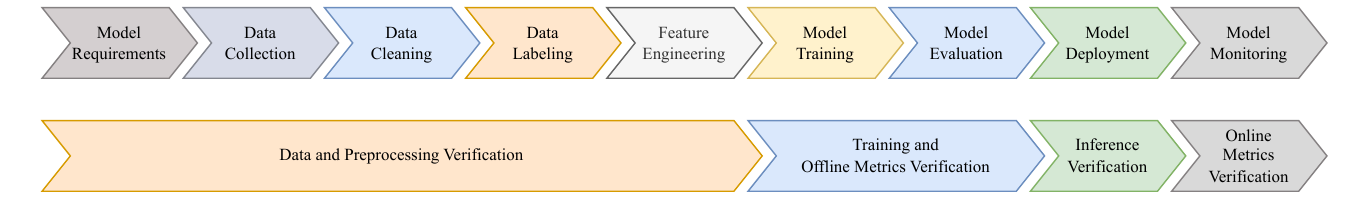}
    \caption{At the top, the diagram depicts the nine phases of the TDSP model~\cite{amershi2019software}, while the bottom illustrates the four phases (grouped) of the MLOps lifecycle verification process derived from the TDSP model.}
    \label{fig:tdsp-cat}
\end{figure*}

The final set of 57 papers was analyzed using a comparative framework designed to highlight key dimensions of existing ZKML approaches:
\begin{itemize}
    \item \emph{ZKP Guarantees}. Completeness, soundness, zero knowledge, and binding properties.
    \item \emph{Adopted Protocols}. Which ZKP protocols were employed.
    \item \emph{Targeted ML Model}. Which ML models were studied for the specific implementation.
    \item \emph{Targeted ML Lifecycle Phase}. Data and Preprocessing Verification, Training and Offline Metrics Verification, Inference Verification, and Online Metrics Verification. These phases are derived through a bucketing process applied to the well-established TDSP model~\cite{amershi2019software}, and a visualization of this process is presented in Figure~\ref{fig:tdsp-cat}. The Data and Preprocessing Verification phase encompasses the verification of properties related to dataset design choices and preprocessing operations. Training and Offline Metrics Verification includes the verification of the training process and the evaluation of model performance using metrics such as accuracy and F1-score, which are computed right after the training. Inference Verification focuses on ensuring the correctness of the inference computation process. Finally, Online Metrics Verification involves the real-time verification of dynamic properties and metrics, such as model drift and live accuracy assessments.
\end{itemize}

The above-mentioned four phases represent the primary aspects currently addressed in the literature concerning the verification of MLOps lifecycle stages. While other established frameworks exist---such as CRISP-DM~\cite{wirth2000crisp} and KDD~\cite{fayyad1996kdd}---the TDSP model was selected for its more fine-grained and comprehensive representation of the MLOps lifecycle. Unlike the aforementioned alternatives, TDSP places less emphasis on business understanding phases, which lie beyond the scope of this work.

This analysis offers a comprehensive overview of the current state of the art in ZKP-enhanced ML, elucidating common challenges and uncovering gaps within the existing literature. Most notably, it reveals a discernible trend toward the convergence of research efforts in this domain, aiming to establish a unified framework for the verification and validation of the overall MLOps lifecycle.

\section{ZKP Protocols Suitability for ML: a Literature Review}
\label{sec:zkp_protocols}
ZKP protocols have evolved into a diverse landscape, with different designs optimized for various computational and security needs. This section categorizes the primary families of ZKP protocols and examines their relevance to ML applications. At the highest level, these protocols can be classified into interactive and non-interactive approaches. Beyond this fundamental distinction, protocols differ in their guarantees, setup requirements, computational representations, post-quantum security, succinctness, and performance characteristics. Each of these factors plays a crucial role in determining a protocol's applicability to verifiable ML. This section provides a structured review of these classification dimensions, highlighting key protocols and their suitability for ML applications. The analysis highlighted seven key dimensions characterizing ZKPs, namely: 
(i) \emph{Interactivity}, 
(ii) \emph{Guarantees Provided by Modern Protocols}, 
(iii) \emph{Setup Requirements}, 
(iv) \emph{Representation of Computation}, 
(v) \emph{Post-Quantum Security Considerations}, 
(vi) \emph{Succinctness Properties}, and 
(vii) \emph{Theoretical Performance Comparison}. 
These properties are further explored in the following sections, and a summary of this analysis on the selected protocols is shown in Table \ref{tab:condensed_protocols}.

\begin{table}
\centering
\tiny
\caption{Condensed Comparison of Cryptographic Protocols.}
\label{tab:condensed_protocols}
\begin{tabular}{lllll}
\toprule
\textbf{Protocol} & \textbf{Interact.} & \textbf{Setup} & \textbf{Post-Quantum Sec.} & \textbf{Succinct.} \\
\midrule
Halo \cite{boweRecursiveProofComposition2019} & Non-Interactive & Universal/Trusted & No & Yes \\
Plonk~\cite{gabizonPLONKPermutationsLagrangebases2019} & Non-Interactive & Universal/Trusted & No & Yes \\
Stark~\cite{ben-sassonScalableTransparentPostquantum2018} & Non-Interactive & Universal/Transparent & Yes & Yes \\
Marlin~\cite{chiesaMarlinPreprocessingZkSNARKs2020} & Non-Interactive & Universal/Trusted & Yes & Yes \\
Sonic~\cite{mallerSonicZeroKnowledgeSNARKs2019} & Non-Interactive & Universal/Trusted & No & Yes \\
Spartan~\cite{settySpartanEfficientGeneralPurpose2020} & Non-Interactive & Universal/Transparent & Yes & Yes \\
Supersonic~\cite{bunzTransparentSNARKsDARK2020} & Non-Interactive & Universal/Transparent & No & Yes \\
Aurora~\cite{ben-sassonAuroraTransparentSuccinct2019} & Non-Interactive & Universal/Transparent & Yes & Yes \\
Fractal~\cite{chiesaFractalPostquantumTransparent2020} & Non-Interactive & Universal/Trusted & Yes & Yes \\
Groth16~\cite{grothSizePairingBasedNoninteractive2016} & Non-Interactive & Universal/Trusted & No & Yes \\
Bulletproofs~\cite{bunzBulletproofsShortProofs2018} & Both & Universal/Transparent & No & Yes \\
Ligero~\cite{amesLigeroLightweightSublinear2017} & Both & Universal/Transparent & Yes & Yes \\
GKR~\cite{goldwasserDelegatingComputationInteractive2015} & Interactive & Universal/Transparent & No & No \\
Wolverine~\cite{wengWolverineFastScalable2021} & Interactive & Universal/Transparent & No & Yes \\
Pinocchio~\cite{parnoPinocchioNearlyPractical2016} & Interactive & Non-Universal/Trusted & No & Yes \\
\bottomrule
\end{tabular}
\end{table}

\subsection{Analysis of Interactivity}

Zero-knowledge protocols can be broadly classified into \emph{interactive} \cite{goldwasser2019knowledge} and \emph{non-interactive} \cite{de1988non} schemes. This distinction directly affects their practicality, particularly in distributed environments or use cases where proofs must be verified repeatedly by independent parties.

Interactive protocols, such as GKR, require a back-and-forth exchange between prover and verifier, where the verifier continuously challenges the prover to validate the computation. While this approach often reduces proof size and prover-side complexity, it requires synchronous communication, limiting scalability in scenarios where proofs are generated once and verified multiple times \cite{jawurek2013zero}.

Non-interactive protocols, including SNARKs, STARKs, etc., compress proof generation into a single exchange, where the prover submits a self-contained proof that any verifier can check independently. This is particularly important in decentralized systems and for applications such as verifiable ML inference, where proofs may be published and validated offline. Non-interactivity in many protocols is achieved via the \emph{Fiat-Shamir heuristic}, which simulates interaction through the use of a hash function acting as a public random oracle \cite{canetti2019fiat}.

\subsection{Guarantees Provided by Modern Protocols}
\label{subsec:guarantees}
All protocols analyzed, spanning interactive, non-interactive, and hybrid approaches, provide the core guarantees defining ZKP protocols: \emph{completeness}, \emph{soundness}, and \emph{zero-knowledge}, as defined by Goldreich et al.\cite{goldreich1994definitions}.

\par 
\emph{Completeness} ensures that a prover following the protocol correctly, with a valid witness, always convinces the verifier. This property is consistently upheld across all surveyed protocols, from early interactive designs to modern non-interactive systems.

\par
\emph{Soundness} guarantees that a dishonest prover, lacking a valid witness, can only convince the verifier with negligible probability. The exact assumptions vary: SNARKs such as Plonk rely on elliptic curve hardness~\cite{gabizonPLONKPermutationsLagrangebases2019}, while hash-based STARKs provide stronger post-quantum resilience~\cite{ben-sassonScalableTransparentPostquantum2018}. Protocols built on Halo inherit soundness from KZG polynomial commitments, similarly tied to elliptic curve assumptions~\cite{boweRecursiveProofComposition2019}.

\par
\emph{Zero-Knowledge} ensures the verifier learns nothing beyond the validity of the claim itself. This is achieved either through blinding techniques in SNARKs \cite{gabizonPLONKPermutationsLagrangebases2019}, or via hash commitments in STARKs~\cite{ben-sassonScalableTransparentPostquantum2018}. In practice, all protocols achieve strong zero-knowledge properties.

A notable point is the frequent use of the \emph{Fiat-Shamir heuristic} \cite{canetti2019fiat} to transform interactive protocols into non-interactive ones, including in Marlin, Spartan. While convenient, this relies on the \emph{Random Oracle Model} (ROM)~\cite{bellare1993random}, weakening formal soundness proofs slightly compared to fully interactive protocols.

Despite minor differences in formalism, all protocols offer guarantees strong enough for real-world privacy-preserving applications~\cite{ernstberger2024you}, including ML inference, provided the chosen protocol aligns with the application’s performance and trust requirements.

\subsection{Setup Requirements}
The setup phase in ZKP systems refers to the preliminary step in which cryptographic parameters are generated before any proving or verification can occur. This phase significantly affects both the security model and the efficiency of the protocol. Broadly, ZKP schemes fall into two categories based on the nature of this setup: those requiring a \emph{trusted setup} and those supporting a \emph{transparent setup} \cite{sheybani2025zero}.

A \emph{trusted setup} involves the generation of a structured reference string (SRS) by a single party or a group of participants. In general, the security assumption hinges on the complete and irreversible disposal of any secret values created during this setup---commonly referred to as \emph{toxic waste}~\cite{jie2019announcing}. If these secrets are ever compromised or retained, an adversary could forge proofs, thus undermining the system’s integrity. While trusted setups can offer compact proofs and fast verification, they introduce a critical vulnerability rooted in the assumption of honest behavior during the setup ceremony.

In contrast, \emph{transparent setups} eliminate the need for trust by deriving public parameters solely from publicly verifiable sources of randomness. Protocols such as zk-STARKs and systems built on Halo exemplify this approach. These protocols do not rely on any secret input during the setup and are therefore inherently more robust in adversarial settings. Transparent setups are particularly appealing for applications requiring strong auditability and long-term trust guarantees, albeit often at the cost of larger proofs and higher prover overhead.

Furthermore, setups can be classified based on their scope as either \emph{universal} or \emph{circuit-specific}. A universal setup, as employed in systems like Marlin and Sonic, supports any computation up to a predefined size and needs to be executed only once. This greatly enhances reusability and reduces setup overhead across multiple applications. On the other hand, circuit-specific setups—as seen in schemes like Pinocchio—require a fresh setup for each distinct computation. While this increases setup cost, it allows for more fine-tuned optimizations tailored to individual circuits.

\subsection{Representation of Computation}
Zero-knowledge protocols do not operate directly on high-level programs or models; instead, they require computations to be transformed into formal representations that are compatible with their internal proof systems~\cite{ernstberger2024you}. These representations play a central role in determining the performance, scalability, and suitability of a protocol for various application domains.

The most widely adopted approach is the \emph{circuit-based representation}, where a computation is expressed as a directed graph: nodes, or \emph{gates}, represent basic operations such as addition or multiplication, and edges, or \emph{wires}, carry intermediate values between operations~\cite{goldwasser2019knowledge}. From a proof system’s perspective, the prover demonstrates knowledge of all wire values — including inputs, outputs, and every intermediate result — and convinces the verifier that these values satisfy the logical constraints imposed by the circuit structure. If any inconsistency is detected, the proof is rejected, ensuring soundness~\cite{goldreich1994definitions}.

Among circuit-based approaches, \emph{arithmetic circuits} are particularly prominent \cite{shpilka2010arithmetic}. These circuits represent computations over finite fields using operations like addition and multiplication. SNARK systems such as Groth16, Plonk, and Marlin operate on a constraint system derived from arithmetic circuits called \emph{Rank-1 Constraint Systems} (R1CS)~\cite{chiesaMarlinPreprocessingZkSNARKs2020}, which translates each gate and wire relationship into a structured set of equations. While efficient for algebraic tasks, arithmetic circuits struggle with non-arithmetic operations---such as comparisons or conditional logic---which must be rewritten or approximated, often adding complexity to the proving process~\cite{jiang2020approximate}.

In contrast, STARKs employ a fundamentally different representation model based on \emph{execution traces}~\cite{ben-sassonScalableTransparentPostquantum2018}. Rather than encoding the computation as a circuit, a STARK captures its dynamic behavior over time. This is done by recording a trace table: a matrix where each row reflects the full state of the computation at a given step, and each column tracks the evolution of a specific variable. This trace is then transformed into an \emph{Algebraic Intermediate Representation} (AIR~\cite{martins2023study}), a set of polynomial constraints that must be satisfied for the trace to be considered valid. While this method offers greater flexibility and post-quantum security, it typically results in larger proofs, particularly for simple or low-complexity programs.

Ultimately, the choice of computational representation shapes not only the cryptographic properties of a proof system but also its practical feasibility for different types of workloads. As such, selecting the appropriate abstraction—be it arithmetic circuits or execution traces—is a critical step in ZKP design.

\subsection{Post-Quantum Security Considerations}
The emergence of quantum computing presents a critical challenge to many cryptographic systems, including a significant subset of ZKP protocols~\cite{chase2017post}. Post-quantum security refers to a protocol’s resistance to adversaries equipped with quantum capabilities — that is, the inability to efficiently break the underlying cryptographic assumptions using quantum algorithms.

Whether a zero-knowledge protocol is considered post-quantum secure depends entirely on the primitives it employs. In general, protocols built solely on \emph{collision-resistant hash functions} (CRHFs \cite{berman2018multi}) are believed to be more resilient in a quantum context, since no quantum algorithm is currently known to break CRHFs faster than brute force. However, it is important to recognize that such protocols are best described as \emph{plausibly} post-quantum secure, as no definitive proof rules out the possibility of future quantum attacks against hash-based constructions~\cite{ben-sassonScalableTransparentPostquantum2018}.

Among the protocols evaluated, \emph{STARKs} are explicitly designed with post-quantum considerations in mind~\cite{ben-sassonScalableTransparentPostquantum2018}. They avoid reliance on number-theoretic assumptions---such as discrete logarithms or elliptic curve pairings---which are known to be vulnerable to quantum attacks like Shor’s algorithm. Instead, STARKs use CRHFs for commitments and integrity checks, making them a compelling choice for applications requiring long-term security and resilience in a post-quantum world.

On the other hand, SNARK-based protocols such as Groth16, Plonk, and Marlin rely on cryptographic assumptions rooted in elliptic curve and pairing-based cryptography~\cite{chiesaMarlinPreprocessingZkSNARKs2020}. These assumptions are susceptible to quantum attacks and therefore cannot be considered post-quantum secure. As such, while these protocols offer strong efficiency and succinctness, they may not be viable for future-proof deployments.

Despite the theoretical urgency, post-quantum security is not yet a central requirement in most current ZKP applications. Nevertheless, as interest grows in areas like secure digital identity, archival data protection, and verifiable computing with long-term guarantees, the demand for cryptographic protocols that can withstand quantum adversaries is expected to rise~\cite{Steinfeld2023Post-Quantum}. Anticipating this shift, future-proof ZKP designs may increasingly favor transparent and hash-based constructions to ensure robust security against emerging threats.

\subsection{Succinctness Properties}
Succinctness is a foundational property of many modern zero-knowledge protocols, particularly those intended for use in bandwidth-limited or resource-constrained environments. A protocol is considered \emph{succinct} if the size of the proof and the time required for its verification scale only polynomially with the size of the input and output, independent of the complexity of the computation being proven~\cite{goldreich1994definitions}. In practice, this means that verification can be performed much faster than re-executing the computation itself, and that the proof remains compact regardless of the underlying workload.

All protocols examined exhibit some form of succinctness, though the degree varies significantly. Classical SNARKs are notable for achieving highly compact proofs---often just a few elliptic curve group elements---and constant-time verification~\cite{chen2022review}. These characteristics make them ideal in scenarios where fast validation and minimal communication overhead are essential. However, their efficiency depends on a trusted setup and cryptographic primitives that are not quantum-resistant.

STARKs, by contrast, are designed for transparency and long-term security~\cite{ben-sassonScalableTransparentPostquantum2018}. They do not require a trusted setup and instead rely on collision-resistant hash functions. While this ensures stronger trust guarantees and potential post-quantum resilience, it leads to considerably larger proofs and longer verification times. This trade-off reflects a shift in priorities, favoring \emph{auditability} and \emph{future-proofing} over minimal proof size.

Protocols based on the GKR framework demonstrate excellent succinctness in individual rounds of interaction, with small messages and lightweight checks~\cite{goldwasserDelegatingComputationInteractive2015}. However, as the number of rounds grows with the depth of the computation, the overall communication and verification costs can accumulate significantly. As a result, while GKR-based approaches are efficient in shallow computations, they may become impractical for deeply nested or complex workloads.

Succinctness, especially in terms of low verification cost, remains a highly desirable property in zero-knowledge systems. It directly impacts the scalability and deployability of these protocols, making them suitable for environments where efficient validation is crucial.



\subsection{Theoretical Performance Comparison}

Zero-knowledge protocols can be broadly evaluated using three core metrics: \emph{prover time}, \emph{verifier time}, and \emph{proof size}~\cite{kobelt2023benchmark}. These theoretical performance estimates, typically expressed in asymptotic terms, offer a first-order approximation of a protocol’s computational efficiency and scalability, independent of implementation details or hardware.


Table~\ref{tab:theoretical_performance} summarizes these asymptotic characteristics for the protocols under consideration. It highlights key distinctions in how each construction handles the burden of proof generation and verification, as well as the cost of communication through proof size.

\begin{table}[ht]
\tiny
\centering
\caption{Theoretical performance of selected zero-knowledge protocols (prover time, verifier time, and proof size).}
\label{tab:theoretical_performance}
\begin{tabular}{@{}llll@{}}
\toprule
\textbf{Protocol} & \textbf{Prover Time} & \textbf{Verifier Time} & \textbf{Proof Size} \\ \midrule
Plonk             & $\mathcal{O}(n\log n)$ & $\mathcal{O}(\log^2 n)$ & $\mathcal{O}(n)$ \\
Marlin            & $\mathcal{O}(n\log n)$ & $\mathcal{O}(|x| + \log n)$ & $\mathcal{O}(n)$ \\
Sonic             & $\mathcal{O}(n\log n)$ & $\mathcal{O}(\log n)$ & $\mathcal{O}(1)$ \\
Spartan           & $\mathcal{O}(n\log n)$ & $\mathcal{O}(n)$ & $\mathcal{O}(\log^2 n)$ \\
Supersonic        & $\mathcal{O}(n\log n)$ & $\mathcal{O}(\log n)$ & $\mathcal{O}(1)$ \\
Stark             & $(\mathcal{O}(n\log n), \mathcal{O}(n^2))$ & $(\mathcal{O}(\log n), \mathcal{O}(n))$ & $(\mathcal{O}(\log n), \mathcal{O}(n))$ \\
Fractal           & $\mathcal{O}(n\log n)$ & $\mathcal{O}(\log^2 n)$ & $\mathcal{O}(\log^2 n)$ \\
Ligero            & $\mathcal{O}(n\log n)$ & $\mathcal{O}(n)$ & $\mathcal{O}(\sqrt{n})$ \\
Aurora            & $\mathcal{O}(n\log n)$ & $\mathcal{O}(n)$ & $\mathcal{O}(\log^2 n)$ \\
Halo              & $\mathcal{O}(n\log n)$ & $\mathcal{O}(\log n)$ & $\mathcal{O}(\log n)$ \\
Bulletproofs      & $\mathcal{O}(n\log n)$ & $\mathcal{O}(n \cdot \log n)$ & $\mathcal{O}(\log n)$ \\
GKR (per round)   & $\mathcal{O}(n^3)$ & $\mathcal{O}(n)$ & $\mathcal{O}(n)$ \\
GKR (overall)     & $\mathcal{O}(n^3\log n)$ & $\mathcal{O}(n\log n)$ & $\mathcal{O}(n\log n)$ \\
Groth16           & $\mathcal{O}(n\log n)$ & $\mathcal{O}(1)$ & $\mathcal{O}(1)$ \\
Pinocchio         & $\mathcal{O}(n\log n)$ & $\mathcal{O}(n)$ & $\mathcal{O}(\log n)$ \\
Wolverine         & NA & NA & NA \\
\bottomrule
\end{tabular}
\end{table}

Among the protocols analyzed, Groth16 is notable for achieving optimal succinctness: it offers constant-size proofs and constant-time verification, making it highly attractive where bandwidth and verifier efficiency are critical. This efficiency, however, comes at the cost of requiring a trusted setup and reliance on elliptic curve pairings~\cite{grothSizePairingBasedNoninteractive2016}.

STARKs, by contrast, avoid any trusted setup and rely solely on collision-resistant hash functions. These choices yield strong transparency and post-quantum security, but result in significantly larger proofs and higher verifier complexity—trade-offs that are intrinsic to their construction~\cite{ben-sassonAuroraTransparentSuccinct2019}.

Other protocols fall along different points in this design space. For instance, systems based on the GKR framework can offer excellent prover efficiency and low communication cost per round, but incur cumulative overhead as the number of rounds grows with the computation’s depth~\cite{goldwasserDelegatingComputationInteractive2015}. Meanwhile, no known protocol achieves prover time better than $\mathcal{O}(n \log n)$, which reflects the additional work required to generate a proof beyond merely executing the underlying computation.

While these theoretical estimates provide useful insights into protocol behavior and scalability, they are not sufficient for drawing conclusions about practical performance. Real-world considerations such as preprocessing costs, memory usage, and parallelization capabilities often play an equally important role. While these aspects are highly relevant to understanding practical performance, they fall outside the scope of this work and should be the focus of future studies, which must include empirical benchmarks and implementation-level evaluations to assess real-world efficiency and scalability.

\subsection{Discussion on ZKP Protocols: Suitability for ML}
\label{sec:suitability_ML}
The application of  ZKPs to ML must span beyond inference alone, extending to training verification, model certification, and integrity assurance across the AI lifecycle. These tasks impose stringent demands on the underlying proof systems, particularly in terms of the guarantees highlighted in Section~\ref{subsec:guarantees}, and compatibility with the structured operations typical of neural networks.

Among the protocol families surveyed, SNARKs and GKR have demonstrated the most practical applicability to ML tasks. SNARKs, such as Groth16 and Plonk, support arithmetic circuits and the Rank-1 Constraint System R1CS format, which aligns well with matrix-based operations in neural networks~\cite{gabizonPLONKPermutationsLagrangebases2019}. Their succinct verification—typically constant-time and constant-size proofs—makes them suitable for low-power or embedded verifiers. However, SNARKs face two main limitations: the reliance on trusted setup ceremonies and the inefficiency in handling non-linear operations, which often require approximations or lookup arguments~\cite{rackoff1991non}.

Recent work has shown that SNARKs can be optimized for ML use through protocol-specific circuit transformations, such as batching matrix operations and reducing the number of constraints~\cite{feng2021zen}. Furthermore, some systems explore \textit{compositional proving}, whereby different ZKPs are combined to prove disjoint parts of a model, each using the most suitable protocol~\cite{campanelliLegoSNARKModularDesign2019}. While prover time remains a challenge, efforts to bring SNARK performance closer to practical deployment continue to advance.

GKR protocols offer a structurally complementary approach, operating directly on layered Boolean circuits, which naturally reflect the feedforward architecture of neural networks~\cite{goldwasserDelegatingComputationInteractive2015}. GKR’s interactive model leads to reduced prover complexity, but requires multiple communication rounds, which can be a limiting factor in asynchronous or decentralized environments. Nonetheless, its low setup requirements and scalable verifier overhead make it well-suited to scenarios where interaction is acceptable or can be transformed into a non-interactive form using the Fiat-Shamir heuristic~\cite{canetti2019fiat}.

STARKs present a compelling alternative due to their transparent setup and post-quantum security. Unlike R1CS-based systems, STARKs use \textit{execution traces} and encode computation through an AIR~\cite{ben-sassonScalableTransparentPostquantum2018}. This enables a broader range of operations but results in significantly larger proofs and longer verification times. Despite these drawbacks, the trend toward quantum-resilient protocols and trust-minimized systems has elevated interest in STARKs for future-proof ZKML deployments.

\begin{figure}
    \centering
    \includegraphics[width=0.8\linewidth]{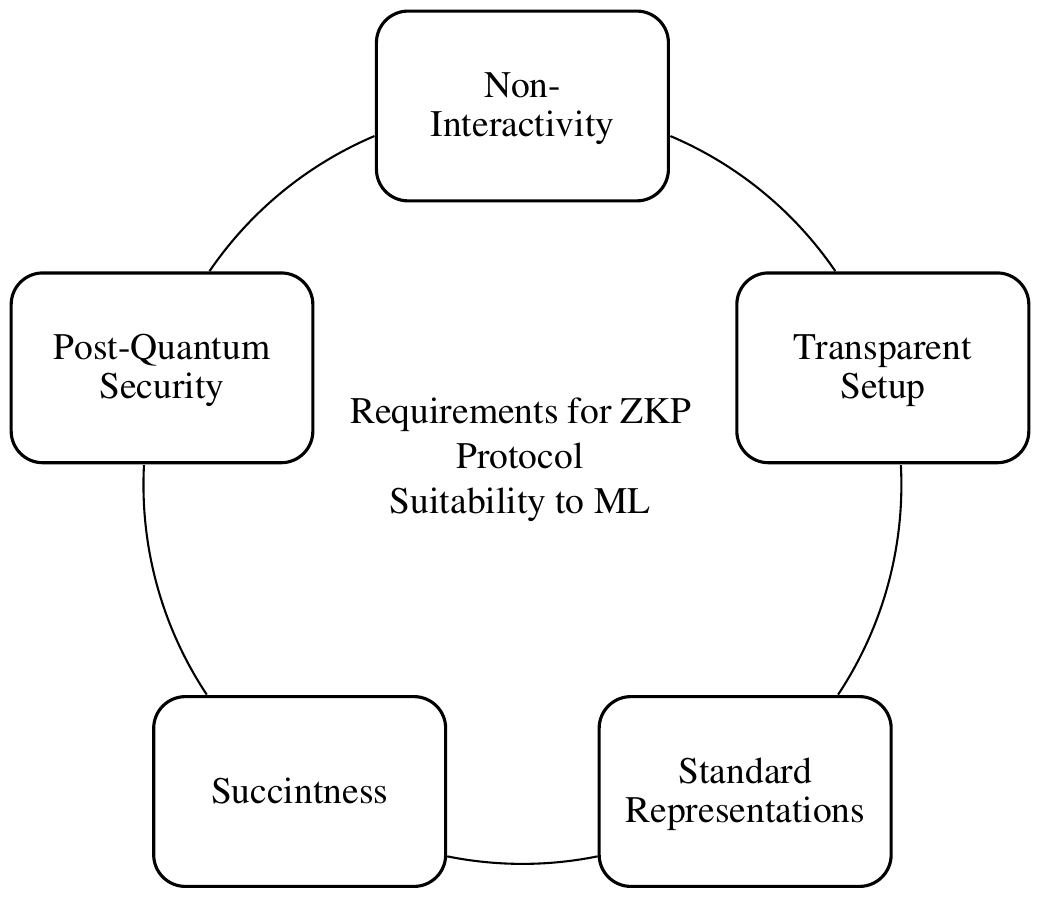}
    \caption{Core properties of ZKP protocols in the context of ML tasks. Each property—ranging from non-interactivity to post-quantum security—reflects emerging trends and practical considerations for deploying ZKPs in real-world ML applications.}
    \label{fig:zkpml_req}
\end{figure}

Here, we outline the 5 key characteristics of a ZKP protocol in the context of ML tasks, highlighting the essential features that enable secure and efficient integration. These properties are outlined in Figure~\ref{fig:zkpml_req}.

\paragraph*{\emph{Non-Interactivity}} While early systems often used interactive protocols, recent trends clearly favor non-interactive designs~\cite{de1988non}. This shift allows a prover to generate a single proof that can be verified by multiple parties without re-execution, significantly reducing overhead in multi-verifier or asynchronous contexts. Many post-2015 protocols adopt the Fiat-Shamir heuristic to transform interactive constructions into non-interactive equivalents~\cite{canetti2019fiat}.

\paragraph*{\emph{Transparent Setup}} As the field matures, transparent setup has emerged as a highly desirable property~\cite{sheybani2025zero}. Protocols that eliminate trusted setup reduce attack vectors and regulatory friction---particularly relevant in medical and financial applications~\cite{Park2023A,Sah2024Efficiency,Kumar2025iZKP}. STARKs and certain variants of Spartan exemplify this direction, using public randomness and hash-based commitments instead of structured reference strings~\cite{settySpartanEfficientGeneralPurpose2020,ben-sassonScalableTransparentPostquantum2018}.

\paragraph*{\emph{Standard Representations}} Most protocols currently rely on circuit-based representations, such as arithmetic circuits or Boolean circuits. R1CS~\cite{lee2021linear} has become a widely adopted standard, particularly within SNARK ecosystems, but it is not universally compatible. STARKs, for instance, use execution traces and AIR, introducing interoperability challenges~\cite{ben-sassonScalableTransparentPostquantum2018}. Having standard and flexible representations is crucial for enabling broader toolchain compatibility, developer accessibility, and seamless integration of ML models into various proof systems.

\paragraph*{\emph{Succinctness}} Succinctness—both in terms of proof size and verifier time—is a near-universal property across modern ZKP systems. This is particularly critical in ZKML, where verifiers may run on constrained hardware, such as mobile devices or edge platforms~\cite{zhong2023demand}. Protocols like Groth16 offer constant-time verification and minimal proof sizes, making them well-suited for scenarios where communication and computational resources are limited~\cite{grothSizePairingBasedNoninteractive2016}.

\paragraph*{\emph{Post-Quantum Security}} Although not yet a baseline requirement in all applications, there is growing awareness of the need for post-quantum secure ZKPs. Protocols such as STARKs, which rely on collision-resistant hash functions rather than elliptic curves or pairings, are well-positioned to address future cryptographic threats~\cite{chiesaFractalPostquantumTransparent2020,Steinfeld2023Post-Quantum}. As quantum-resistant infrastructure becomes more pressing, support for this property may become critical.

Despite these promising trends, several challenges remain. The most significant is the performance, which, under all current constructions, remains bounded below by \(\mathcal{O}(n \log n)\) (see Table~\ref{tab:theoretical_performance}). Furthermore, in practice, ZKP implementations often suffer from significant constant overheads introduced by compiler inefficiencies, memory consumption, and limited backend parallelism~\cite{Samudrala2024Performance,2023GZKP}. 

\section{ZKP-Enhanced ML: a Literature Review}
\label{sec:zkml}

This section presents a systematic review of the existing research landscape on ZKP-Enhanced ML, also known as Zero-Knowledge Machine Learning (ZKML), identifying key approaches and methodologies employed to construct ZKPs for ML applications. The analysis focuses on how different works address efficiency bottlenecks, optimize proof generation, and manage trade-offs between proof succinctness and computational overhead. By examining the evolution of these methods in chronological order, this review highlights the current state of the art, revealing emerging patterns and the convergence of the research domain toward a unified ZKMLOps framework for Trustworthy ML development.

\subsection{Overview of Existing Research}
The solutions presented in existing research address several ML-related topics, which can be broadly grouped into two main types of contributions: \emph{Federated Learning} (FL) and \emph{ML as a Service} (MLaaS). We identified 26 papers focusing on FL (based on the definition by Bonawitz et al.\cite{bonawitz2017practical}) and 31 papers on MLaaS (based on the definition by Hesamifard et al.\cite{hesamifard2017privacy}).
The 26 papers addressing FL primarily study problems related to the privacy and confidentiality of user data, the integrity of aggregation processes, and local updates to prevent poisoning attacks. Among these, 16 papers adopt techniques of verifiable computing, such as homomorphic encryption (e.g.,\cite{guoVeriFLCommunicationEfficientFast2021, madiSecureFederatedLearning2021}), differential privacy (e.g.,\cite{chengUniversalInteractiveVerification2022}), or chain mechanisms (e.g.,~\cite{pengVFChainEnablingVerifiable2022}). The remaining 10 FL papers employ ZKP techniques. As further exploration of these FL studies is planned for future work, they are not analyzed in detail here. The list of these papers can still be found in the replication package mentioned in Section~\ref{sec:methodology}.
With respect to the 30 papers addressing MLaaS, on which we focused our analysis, the goals typically revolve around guaranteeing: (i) integrity of the computation, (ii) privacy and confidentiality, and (iii) fairness between parties. Of these, 13 papers apply techniques such as homomorphic encryption (e.g.,\cite{froelicherDrynxDecentralizedSecure2020,liVerifiablePrivacyPreservingMachine2022,xuSecureVerifiableInference2020}), randomized algorithms (e.g.,\cite{huangZkMLaaSVerifiableScheme2022,shaoPrivacyPreservingVerifiableCloudAided2022,zhangHydraPipelineableInteractive2021}), or blockchains (e.g.,\cite{aleisaTAIRABSCTrustingAI2022,ramanScalableBlockchainApproach2019,zhangMachineLearningCloud2022}). Our analysis focuses on the remaining 17 MLaaS papers that employ ZKP techniques or provide new ZKP implementations for ML applications:\cite{abbaszadehZeroKnowledgeProofsTraining2024,chenZKMLOptimizingSystem2024,fengZENOptimizingCompiler2021,fengZENOTypebasedOptimization2024,ghaffaripourMutuallyPrivateVerifiable2021,juEfficientSumCheckProtocol2021,leeVCNNVerifiableConvolutional2024a,liuZkCNNZeroKnowledge2021a,luEfficientExtensibleZeroknowledge2024,sunZkDLEfficientZeroKnowledge2025,sunZkLLMZeroKnowledge2024,toreiniFairnessServiceFaaS2024,waiwitlikhitTrustlessAuditsRevealing2024,wangEfficientZeroKnowledgeClassical2025,wuConfidentialVerifiableMachine2024,zhangZeroKnowledgeProofs2020a,zhaoVeriMLEnablingIntegrity2021}. These contributions will be further discussed in the following section.

\subsection{Analysis of the ZKML Approaches}
This section will provide a concise summary of the approaches identified in the literature. This comprehensive analysis is essential, as the proposed approaches address distinct aspects and propose varying solutions to the challenges they seek to overcome. Furthermore, these challenges exhibit significant variability. 

Zhang et al.~\cite{zhangZeroKnowledgeProofs2020a} initiated the exploration of ZKPs in the context of ML tasks, with a focus on verifying both predictions and model accuracy. They proposed an efficient scheme tailored to zero-knowledge decision trees. Specifically, their contributions include: (i) the design of an efficient protocol for ZKPs of decision tree predictions; (ii) the extension of this protocol to support accuracy verification of decision trees in zero knowledge, incorporating task-specific optimizations; and (iii) the implementation and empirical evaluation of the proposed protocol. The underlying proof system utilized is Aurora~\cite{ben2019aurora}. We further categorized this work under \emph{Inference Verification} and \emph{Online Metrics Verification}.

Liu et al.~\cite{liu2021zkcnn} propose an efficient ZKP scheme for CNN predictions and accuracy that scales to large CNN models, enabling the computation of such proofs without the excessive overhead introduced by general-purpose ZKP schemes that work for any computations modeled as arithmetic circuits. This improvement is based on a novel sum-check protocol based on the \emph{Fast Fourier Transform} (FFT). The proposed scheme is then extended, adding generalization and integration with the GKR protocol~\cite{goldwasserDelegatingComputationInteractive2015}. We further categorized this work under \emph{Inference Verification} and \emph{Online Metrics Verification}.

Ju et al.~\cite{juEfficientSumCheckProtocol2021} propose a new efficient sum-check protocol for a CNN convolution operation, achieving an asymptotically optimal proving cost for a convolution operation. Their scheme employs a combination of the sum-check protocol~\cite{lund1992algebraic}, and GKR~\cite{goldwasserDelegatingComputationInteractive2015}. The protocol is then evaluated, and it is shown how it improves previous work on verifiable CNNs~\cite{liu2021zkcnn} reaching optimal computation cost and smaller proof size. We further categorized this work under \emph{Inference Verification}.

Ghaffaripour et al.~\cite{ghaffaripourMutuallyPrivateVerifiable2021} address the challenge of assuring the integrity of computations performed by MLaaS platforms, by proposing a novel distributed approach which uses specialized composable proof systems at its core. More precisely, the mathematical formulation of the ML task is divided into multiple parts, each of which is handled by a different specialized proof system; these proof systems are then combined with the commit-and-prove methodology to guarantee correctness as a whole. This methodology is based on the implementation of LegoSNARK~\cite{campanelliLegoSNARKModularDesign2019}, a toolbox for \emph{commit-and-prove zkSNARKs} (CP-SNARKs). The solution is evaluated against a verification of the integrity of a classification task on a \emph{Support Vector Machine}. We further categorized this work under \emph{Inference Verification}.

Zhao et al.~\cite{zhaoVeriMLEnablingIntegrity2021} propose VeriML, a MLaaS framework that provides tunable probabilistic assurance on service correctness as well as service fee accounting fairness. To achieve this, VeriML utilizes a novel CP-SNARK protocol on randomly selected iterations during the ML training phase. Moreover, in doing so, it utilizes multiple circuit-friendly optimizations for the verification of expensive operations such as matrix multiplication and non-linear functions in ML algorithms. The authors empirically validate the efficiency of the proposed solutions on several ML models, namely linear regression, logistic regression, neural network, support vector machines, K-Means, and decision tree. We further categorized this work under \emph{Training and Offline Metrics Verification} and \emph{Inference Verification}.

Feng et al.~\cite{feng2021zen} present ZEN, the first attempt in the literature to provide an optimizing compiler that generates efficient verifiable, zero-knowledge neural network accuracy ($\text{ZEN}_{acc}$) and inference ($\text{ZEN}_{infer}$) schemes. The first is used to verify that a committed neural network model achieves a claimed accuracy on a test dataset without revealing the model itself. The latter, instead, is used to verify that the inference result from the private model on a given input is correct, without revealing the model or the input. Since the direct application of pure zkSNARKs for these tasks requires prohibitive computational costs, the authors first incorporate a new neural network quantization algorithm that incorporate two R1CS friendly optimizations which makes the model to be express in zkSNARKs with less constraints and minimal accuracy loss; second, ZEN introduces a SIMD style optimization, namely stranded encoding, that can encode multiple 8bit integers in large finite field elements without overwhelming extraction cost. We further classified this work under offline metrics verification and \emph{Inference Verification}.

Garg et al.~\cite{garg2022succinct} propose a novel method for verifying floating-point computations that guarantees approximate correctness w.r.t. a relative error bound. The standard approach to handling floating-point computations requires conversion to binary circuits, following the IEEE-754 floating-point standard. This approach incurs a $\text{poly}(w)$ overhead in prover efficiency for computations with $w$-bit precision, resulting in very high prover runtimes, which is still one of the main issues and bottlenecks in the design of succinct arguments. The proposed solution consists of a compiler optimization that incurs only a $\log(w)$ overhead in the prover's running time. Although this work does not provide a proving scheme tailored specifically for ML tasks, it paves the way for further research in ML and scientific computing by providing an efficient way of proving possibly any ML-pipeline phase that involves floating-point computations.

Toreini et al.~\cite{toreiniFairnessServiceFaaS2024} propose FaaS, an auditing framework that emphasizes trustworthy AI, particularly group fairness. Group fairness refers to the property that the demographics of individuals receiving positive (or negative) classifications are consistent with the demographics of the entire population~\cite{dwork2012fairness}. In other words, an ML model is considered fair (in the context of group fairness) if it treats different groups equally~\cite{mehrabi2021survey}. In particular, FaaS is a privacy-preserving, end-to-end verifiable architecture to collectively audit the algorithmic fairness of ML systems. FaaS is model-agnostic (independent of the ML model) and takes a holistic approach towards auditing for group fairness metric. More precisely, the authors propose an auditing approach based on a 1-out-of-n interactive ZKP technique, famously known as CDS (Cramer, Damgard, and Schoenmakers)~\cite{cramer1994proofs,cramer1997secure}. Although promising, the solution is based on the strong assumption that the ML system presents the data and predictions honestly. We further classified the work under \emph{Online Metrics Verification}.

Feng et al.~\cite{fengZENOTypebasedOptimization2024} present \textsc{ZENO} (ZEro-knowledge Neural network Optimizer), a type-based optimization framework designed to enable efficient neural network inference verification. In conventional zkSNARK systems~\cite{feng2021zen}, arbitrary arithmetic functions are compiled into low-level arithmetic circuits, thereby discarding high-level neural network semantics such as tensor structure and privacy guarantees, which become difficult to reconstruct. The authors address this limitation as their first contribution by proposing a novel language construct that preserves high-level semantics throughout zkSNARK proof generation. Their second contribution introduces an optimized circuit generation strategy that leverages this preserved semantic information to reduce both computational complexity and the total number of operations. The third contribution consists of a neural network-centric system-level optimization that further enhances the performance of zkSNARKs when applied to neural network inference tasks. The framework is implemented atop general-purpose zkSNARK methodologies and benchmarked against existing tools following a similar design philosophy, including Arkworks~\cite{ArkworksrsSnark2025}, Bellman~\cite{ZkcryptoBellman2025}, and Ginger~\cite{HorizenOfficialGingerlib2024}. We categorize this work under \emph{Inference Verification}.

\par
Chen et al.~\cite{chenZKMLOptimizingSystem2024} introduce \textsc{ZKML}, a framework designed to generate zkSNARKs~\cite{ben-sassonScalableTransparentPostquantum2018} for realistic and complex ML models. This work specifically targets the \texttt{halo2} proving system~\cite{ZcashHalo22025}, which incorporates the Plonkish randomized AIR (Arithmetic Intermediate Representation) with preprocessing~\cite{gabizon2021airs}. The framework represents a significant advancement, enabling the computation of zkSNARKs for a diverse set of models with realistic scales and structures for the first time. The authors demonstrate the capabilities of \textsc{ZKML} by applying it to several representative models, including a distilled version of GPT-2 (81.3M parameters), a diffusion model (19.4M parameters), Twitter's recommender system (48.1M parameters), DLRM (764.3K parameters), MobileNet (3.5M parameters), ResNet-18 (280.9K parameters), VGG16 (15.2M parameters), and MNIST (8.1K parameters). This contribution is further categorized under \emph{Inference Verification}.

\par
Sun et al.~\cite{sunZkLLMZeroKnowledge2024} propose a specialized ZKP framework tailored to Large Language Models (LLMs). Their work introduces two key components: \texttt{tlookup}, a ZKP protocol designed to support universal non-arithmetic operations commonly encountered in deep learning; and \texttt{zkAttn}, a ZKP protocol specifically crafted to verify attention mechanisms in LLMs. The \texttt{zkAttn} protocol is built upon the sumcheck protocol~\cite{bootle2021sumcheck} and the Hyrax protocol~\cite{wahby2018doubly}, ensuring efficient and scalable proof generation for the attention layer. The proposed framework is evaluated on prominent LLM architectures, including OPT and LLaMa-2. This contribution is further categorized under \emph{Inference Verification}.

\par
Sun et al.~\cite{sunZkDLEfficientZeroKnowledge2025} present \textsc{zkDL}, an efficient ZKP framework for deep learning training. To enhance performance, the authors introduce \texttt{zkReLU}, a specialized ZKP protocol optimized for the exact computation of the ReLU activation function and its backpropagation. Furthermore, the authors propose \textsc{FAC4DNN}, a modeling scheme that captures the training process of deep neural networks using arithmetic circuits grounded in the GKR protocol~\cite{goldwasserDelegatingComputationInteractive2015}. The framework is empirically evaluated on an 8-layer neural network comprising over 10 million parameters. This contribution is categorized under \emph{Training and Offline Metrics Verification}.

\par
Wu et al.~\cite{wuConfidentialVerifiableMachine2024} present a confidential and verifiable delegation scheme for ML inference in untrusted cloud environments. Their work focuses on enabling both privacy and integrity by combining secure multiparty computation with ZKPs. The core of their approach uses interactive proofs, specifically, the GKR~\cite{goldwasserDelegatingComputationInteractive2015} protocol enhanced with polynomial commitments, to generate efficient, low-overhead proofs, even when most of the participating servers are potentially malicious. The protocol is optimized for arithmetic circuits and includes a custom design for matrix multiplication that significantly reduces proof generation time. Experimental results on neural networks, including a 3-layer fully connected model and LeNet, show large performance gains compared to prior work. We classify this contribution under \emph{Inference Verification}.

\par
Lee et al.~\cite{leeVCNNVerifiableConvolutional2024a} introduce vCNN, a verifiable convolutional neural network framework that addresses the inefficiency of zk-SNARK-based inference verification for CNNs. Their key innovation lies in optimizing the representation of convolutional operations, which dominate CNN computations, by proposing a novel QPP-based formulation that reduces proving complexity from $O(ln)$ to $O(l + n)$. To handle other network components such as ReLU and pooling, which are not efficiently supported by QPP, they combine QPP and QAP circuits and use CP- and cc-SNARKs~\cite{campanelliLegoSNARKModularDesign2019} to link them, enabling efficient end-to-end proof generation. Their model supports standard CNNs like MNIST, AlexNet, and VGG16, achieving up to 18{,}000$\times$ speedups in proof generation time and drastic reductions in CRS size compared to prior zk-SNARK approaches~\cite{grothSizePairingBasedNoninteractive2016,parnoPinocchioNearlyPractical2016}. We classify this work under \emph{Inference Verification}.

\par  
Abbaszadeh et al.~\cite{abbaszadehZeroKnowledgeProofsTraining2024} propose Kaizen, a ZKP of training (zkPoT) system designed for deep neural networks. The goal is to enable a party to prove that a model was correctly trained on a committed dataset using gradient descent, without revealing either the model or the data. Their construction combines an optimized GKR-style proof system~\cite{goldwasserDelegatingComputationInteractive2015} for single gradient descent steps with a recursive composition framework to achieve succinctness across multiple iterations. A novel contribution is their aggregatable polynomial commitment scheme tailored for multivariate polynomials, which is essential for scaling recursive proofs efficiently. Kaizen supports large models like VGG-11 and demonstrates a prover time of 15 minutes per iteration, 24× faster and 27× more memory-efficient than generic recursive ZK schemes, with proof size and verifier time independent of iteration count. We classify this work under data and \emph{Training and Offline Metrics Verification}.

\par
Wang et al.~\cite{wangEfficientZeroKnowledgeClassical2025} propose ezDPS, a zero-knowledge framework for verifying classical ML inference pipelines in outsourced settings. The pipeline comprises four stages: data denoising using Discrete Wavelet Transform, normalization with Z-Score, feature extraction via Principal Component Analysis, and classification using Support Vector Machines. Each stage is converted into arithmetic circuits using custom-designed zero-knowledge gadgets for core operations, including square root, exponentiation, max/min, and absolute value. The framework is instantiated over the Spartan CP-ZKP backend~\cite{settySpartanEfficientGeneralPurpose2020}, supporting efficient Rank-1 Constraint Systems with polynomial commitments. ezDPS introduces a zkPoA (zero-knowledge Proof-of-Accuracy) scheme, allowing the server to prove that a committed model achieves a specified minimum accuracy over public datasets without revealing model parameters. To improve efficiency, the authors leverage techniques like random linear combination for dimensionality reduction and permutation-based maximum value selection. We classify this work under \emph{Data and Preprocessing Verification}, \emph{Inference Verification}, and \emph{Online Metrics Verification}.

Waiwitlikhit et al.~\cite{waiwitlikhitTrustlessAuditsRevealing2024} propose \textsc{ZKAUDIT}, a zero-knowledge audit framework enabling trustless verification of model training and data properties without revealing model weights or training data. The system consists of two main phases: \textsc{ZKAUDIT-T}, which proves that a model was trained via stochastic gradient descent on a committed dataset, and \textsc{ZKAUDIT-I}, which allows auditing arbitrary properties over the hidden data and weights through user-defined functions. The framework leverages ZK-SNARKs over AIRs, using the Halo2~\cite{ZcashHalo22025} backend with optimizations such as rounded division, variable fixed-point precision, and softmax implementation in finite fields. It supports real-world models like MobileNet v2 and DLRM-style recommenders. The framework supports audits such as censorship detection, copyright verification, and counterfactual analysis. We classify this work under \emph{Data and Preprocessing Verification}, and \emph{Training and Offline Metrics Verification}.

\subsection{Discussion on ZKP-Enhanced ML: An MLOps Lifecycle Overview}

This section presents a discussion of the primary findings from the survey on ZKP-Enhanced ML applications, with an emphasis on the MLOps verification lifecycle inspired by the TDSP model~\cite{amershi2019software}, introduced in Section~\ref{sec:methodology} and Figure~\ref{fig:tdsp-cat}. To structure this analysis, we divide the discussion into two phases. In the first phase, we describe the main findings by identifying the specific phase of the MLOps verification lifecycle addressed in each work, the model used, and the protocol employed—this latter aspect being assessed through the ZKP-ML suitability model defined in Section~\ref{sec:suitability_ML}. The second phase of the analysis highlights a central insight of our investigation: the identification of a convergence trend across the reviewed literature, pointing toward the development of a unified and comprehensive model for MLOps verification in the broader context of Trustworthy AI.

\subsubsection{MLOps Verification Lifecycle: Phases, Models and Protocols}
\begin{figure}
\begin{tikzpicture}
\node[anchor=south west,inner sep=0] at (0,0) {\includegraphics[width=1\linewidth]{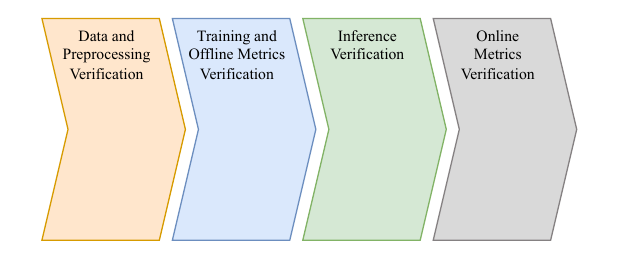}};
\draw (1.6,2.2) node{{\scriptsize 
\cite{wangEfficientZeroKnowledgeClassical2025,waiwitlikhitTrustlessAuditsRevealing2024}}};
\draw (3.5, 2.2) node{{\scriptsize \cite{zhaoVeriMLEnablingIntegrity2021,sunZkDLEfficientZeroKnowledge2025},}};
\draw (3.5, 1.9) node{{\scriptsize \cite{abbaszadehZeroKnowledgeProofsTraining2024,waiwitlikhitTrustlessAuditsRevealing2024},}};
\draw (5.4, 2.2) node{{\scriptsize \cite{zhangZeroKnowledgeProofs2020a,liu2021zkcnn},}};
\draw (5.4, 1.9) node{{\scriptsize \cite{juEfficientSumCheckProtocol2021,ghaffaripourMutuallyPrivateVerifiable2021},}};
\draw (5.35, 1.6) node{{\scriptsize \cite{zhaoVeriMLEnablingIntegrity2021,feng2021zen},}};
\draw (5.4, 1.3) node{{\scriptsize \cite{fengZENOTypebasedOptimization2024,chenZKMLOptimizingSystem2024},}};
\draw (5.4, 1) node{{\scriptsize \cite{sunZkLLMZeroKnowledge2024,wuConfidentialVerifiableMachine2024},}};
\draw (5.37, 0.7) node{{\scriptsize \cite{leeVCNNVerifiableConvolutional2024a,wangEfficientZeroKnowledgeClassical2025}}};
\draw (7.3, 2.2) node{{\scriptsize \cite{zhangZeroKnowledgeProofs2020a,liu2021zkcnn},}};
\draw (7.3, 1.9) node{{\scriptsize \cite{toreiniFairnessServiceFaaS2024,wangEfficientZeroKnowledgeClassical2025},}};
\end{tikzpicture}
\caption{ZKP-Enhanced ML applications in the MLOps verification lifecycle.}
\label{fig:zkml-classification}
\end{figure}
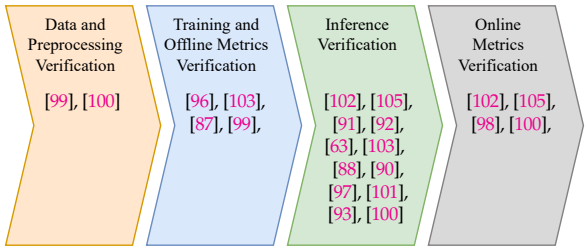

In our survey and classification of the literature, we identified a diverse range of efforts addressing different stages of the MLOps verification lifecycle. This classification can be seen in Figure~\ref{fig:zkml-classification}. Specifically, we observed that two studies explicitly target the phase of \emph{Data and Preprocessing Verification}, four contributions focus on \emph{Training and Offline Metrics Verification}, a significantly larger group of twelve papers address \emph{Inference Verification}, and four works propose solutions for \emph{Online Metrics Verification}. This distribution of research efforts highlights a substantial emphasis on the inference stage, suggesting that the research community currently prioritizes the integrity and correctness of model predictions during deployment. This trend is perhaps unsurprising, as the inference phase is typically the most security-sensitive and externally exposed component of the ML lifecycle in real-world deployments. It also presents some of the most significant technical challenges, particularly in the efficient generation and verification of ZKPs. These challenges have made inference the primary focus of recent research, as it represents the most prominent bottleneck in achieving practical, verifiable ML systems.

However, this imbalance also reveals notable research gaps. In particular, comparatively limited attention has been paid to the earlier stages of the pipeline, such as data acquisition, preprocessing, and training integrity. These stages are no less important: they are foundational to model correctness, fairness, and generalization, and can often be the origin of subtle but critical vulnerabilities or data misuse. Encouragingly, some recent works have started to adopt a more holistic view, proposing solutions that span multiple verification phases or that attempt to encompass the entire ML lifecycle ZKP frameworks~\cite{wang2024efficient}. This evolving trend toward end-to-end verifiability is a promising direction for future work.

\begin{table}
\centering
\scriptsize
\caption{ML Models Studied in the ZK-Enhanced ML Literature.}
\label{tab:ml-models}
\begin{tabular}{@{}ll@{}}
\toprule
\textbf{ML Model Category} & \textbf{References} \\ \midrule

Decision Trees & 
\cite{zhangZeroKnowledgeProofs2020a}, \cite{zhaoVeriMLEnablingIntegrity2021} \\

Support Vector Machines & 
\cite{ghaffaripourMutuallyPrivateVerifiable2021}, \cite{zhaoVeriMLEnablingIntegrity2021}, \cite{wangEfficientZeroKnowledgeClassical2025} \\

Linear Models (Linear/Logistic Regression) & 
\cite{zhaoVeriMLEnablingIntegrity2021} \\

Clustering (K-Means) & 
\cite{zhaoVeriMLEnablingIntegrity2021} \\

General Neural Networks & 
\cite{feng2021zen}, \cite{fengZENOTypebasedOptimization2024}, \cite{chenZKMLOptimizingSystem2024}, 
\cite{zhaoVeriMLEnablingIntegrity2021}, \cite{wuConfidentialVerifiableMachine2024}, \cite{sunZkDLEfficientZeroKnowledge2025} \\

Convolutional Neural Networks & 
\cite{liu2021zkcnn}, \cite{juEfficientSumCheckProtocol2021}, \cite{leeVCNNVerifiableConvolutional2024a} \\

Large Language Models & 
\cite{sunZkLLMZeroKnowledge2024,chenZKMLOptimizingSystem2024} \\

Vision Models (VGG, ResNet, MobileNet, LeNet) & 
\cite{chenZKMLOptimizingSystem2024}, \cite{leeVCNNVerifiableConvolutional2024a}, 
\cite{abbaszadehZeroKnowledgeProofsTraining2024}, \cite{wuConfidentialVerifiableMachine2024}, 
\cite{waiwitlikhitTrustlessAuditsRevealing2024} \\

Recommender Systems (DLRM, Twitter) & 
\cite{chenZKMLOptimizingSystem2024}, \cite{waiwitlikhitTrustlessAuditsRevealing2024} \\

\bottomrule
\end{tabular}
\end{table}

In terms of the types of ML models addressed by the reviewed literature, Table~\ref{tab:ml-models} provides a summary of the distribution across model classes. A clear trend emerges in favor of complex deep learning models, particularly \emph{Neural Networks} and \emph{Convolutional Neural Networks}, which have become dominant in both academic research and real-world applications due to their high expressive power and state-of-the-art performance across many domains. This focus aligns with the technical challenges posed by these models, such as large parameter counts, non-linear activations, and costly inference operations, which make their verification particularly demanding and thus an attractive target for ZKP-based approaches.

Nevertheless, it is worth noting that several contributions also address traditional ML models, including \emph{Decision Trees}, \emph{Support Vector Machines}, \emph{Linear and Logistic Regression}, and \emph{Clustering algorithms} like \emph{K-Means}. These classical models remain widely used in industry due to their interpretability, efficiency, and performance in low-data regimes. The presence of works tackling these models demonstrates a healthy diversity in research, and it is especially encouraging as these simpler models can serve as testbeds for novel ZKP constructions or optimizations that may later be scaled to more complex architectures.

\begin{figure}
    \centering
    \includegraphics[width=\linewidth]{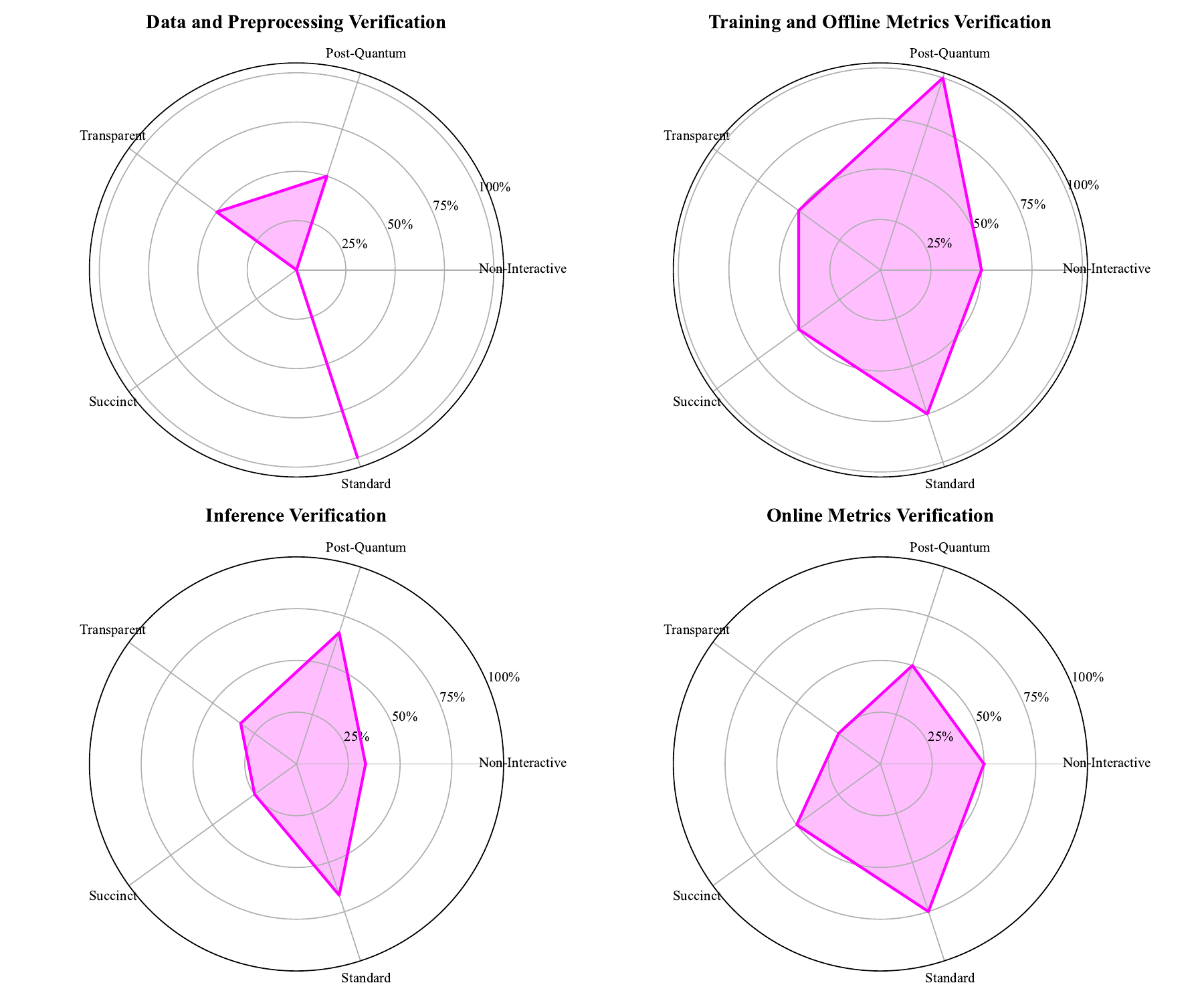}
    \caption{ZKP Protocols suitability to ML Applications for every MLOps Verification phase.}
    \label{fig:zkml-suitability}
\end{figure}

Turning to the analysis of ZKP protocol suitability, we evaluated the extent to which the underlying cryptographic protocols used in each work satisfy the key properties required for practical integration in ML workflows as described in Section~\ref{sec:suitability_ML}.
Figure~\ref{fig:zkml-suitability} summarizes the degree to which current works meet these criteria across the defined MLOps phase. None of the surveyed phases exhibit full compliance with these properties across all works. Across all phases, at least some of the reviewed works rely on cryptographic protocols that do not fully adhere to our defined suitability criteria. These shortcomings highlight that, despite meaningful progress in recent years, substantial effort is still required to design and standardize ZKP systems that are not only theoretically robust but also practically viable for integration into contemporary ML pipelines.

\subsubsection{Convergence Towards a Unified MLOps Verification Model}
After analyzing how zero-knowledge protocols are applied across the MLOps verification lifecycle, we observed a convergence of efforts toward a unified framework for Trustworthy AI, which we term \emph{ZKMLOps}. This framework integrates ZKPs into ML pipelines to provide strong cryptographic guarantees of correctness, integrity, and privacy. We categorized existing work into three classes: \emph{Enabling Technologies}, \emph{Applied Verification}, and \emph{Trustworthy AI}.

While the majority of contributions fall within the first two categories, only a few works---Toreini et al.~\cite{toreiniFairnessServiceFaaS2024} and Waiwitlikhit et al.~\cite{waiwitlikhitTrustlessAuditsRevealing2024}—explicitly address core trustworthy AI principles such as fairness, copyrights, censorship, and counterfactual audits. Nonetheless, this should not be seen as a limitation. The inherent properties of ZKPs are naturally aligned with key trustworthy AI goals, including privacy and data governance, accountability and auditability, and transparency~\cite{thiebesTrustworthyArtificialIntelligence2021,kaurTrustworthyArtificialIntelligence2022a}.

\begin{figure}
    \begin{tikzpicture}
    \node[anchor=south west,inner sep=0] at (0,0) {\includegraphics[width=1\linewidth]
    {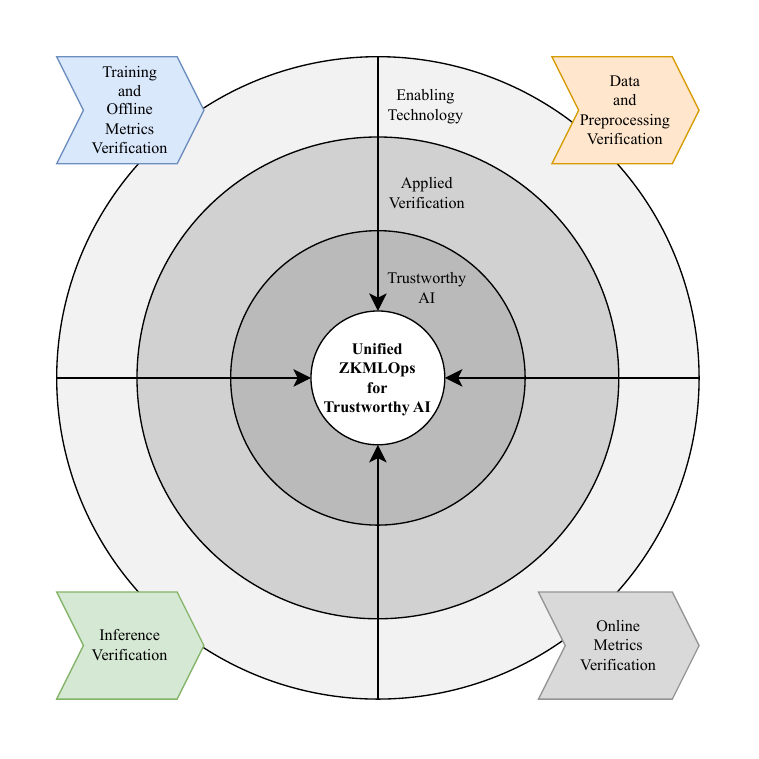}};

    \draw (6,6.2) node{{\scriptsize \cite{wangEfficientZeroKnowledgeClassical2025}}};
    \draw (6.05,5.9) node{{\scriptsize (2025)}};
    \draw (6.4,5.4) node{{\scriptsize \cite{waiwitlikhitTrustlessAuditsRevealing2024}}};
    \draw (6.45,5.1) node{{\scriptsize (2024)}};
    \draw (5.5,5.1) node{{\scriptsize \cite{waiwitlikhitTrustlessAuditsRevealing2024}}};
    \draw (5.55,4.8) node{{\scriptsize (2024)}};

    \draw (1.5,6.2) node{{\scriptsize \cite{sunZkDLEfficientZeroKnowledge2025}}};
    \draw (1.55,5.9) node{{\scriptsize (2025)}};
    \draw (2.2,5.4) node{{\scriptsize \cite{zhaoVeriMLEnablingIntegrity2021}}};
    \draw (2.25,5.1) node{{\scriptsize (2021)}};
    \draw (2.7,6.2) node{{\scriptsize \cite{abbaszadehZeroKnowledgeProofsTraining2024}}};
    \draw (2.75,5.9) node{{\scriptsize (2024)}};
    \draw (3.6,6.7) node{{\scriptsize \cite{waiwitlikhitTrustlessAuditsRevealing2024}}};
    \draw (3.65,6.4) node{{\scriptsize (2024)}};
    \draw (3.6,5.6) node{{\scriptsize \cite{waiwitlikhitTrustlessAuditsRevealing2024}}};
    \draw (3.65,5.3) node{{\scriptsize (2024)}};

    \draw (1.2,4) node{{\scriptsize \cite{juEfficientSumCheckProtocol2021}}};
    \draw (1.25,3.7) node{{\scriptsize (2021)}};
    \draw (1.4,3.2) node{{\scriptsize \cite{fengZENOTypebasedOptimization2024}}};
    \draw (1.45,2.9) node{{\scriptsize (2024)}};
    \draw (1.9,2.4) node{{\scriptsize \cite{chenZKMLOptimizingSystem2024}}};
    \draw (1.95,2.1) node{{\scriptsize (2024)}};
    \draw (2.8,1.7) node{{\scriptsize \cite{sunZkLLMZeroKnowledge2024}}};
    \draw (2.85,1.4) node{{\scriptsize (2024)}};
    \draw (3.7,1.4) node{{\scriptsize \cite{garg2022succinct}}};
    \draw (3.75,1.1) node{{\scriptsize (2022)}};

    \draw (2.1,4.2) node{{\scriptsize \cite{zhangZeroKnowledgeProofs2020a}}};
    \draw (2.15,3.9) node{{\scriptsize (2020)}};
    \draw (2,3.6) node{{\scriptsize \cite{liu2021zkcnn}}};
    \draw (2.05,3.3) node{{\scriptsize (2021)}};
    \draw (2.7,3.6) node{{\scriptsize \cite{zhaoVeriMLEnablingIntegrity2021}}};
    \draw (2.75,3.3) node{{\scriptsize (2021)}};
    \draw (2.4,3) node{{\scriptsize \cite{ghaffaripourMutuallyPrivateVerifiable2021}}};
    \draw (2.45,2.7) node{{\scriptsize (2021)}};
    \draw (3.1,3) node{{\scriptsize \cite{feng2021zen}}};
    \draw (3.15,2.7) node{{\scriptsize (2021)}};
    \draw (3.1,2.4) node{{\scriptsize \cite{wuConfidentialVerifiableMachine2024}}};
    \draw (3.15,2.1) node{{\scriptsize (2024)}};
    \draw (3.9,2.7) node{{\scriptsize \cite{leeVCNNVerifiableConvolutional2024a}}};
    \draw (3.95,2.4) node{{\scriptsize (2024)}};
    \draw (4,2.1) node{{\scriptsize \cite{wangEfficientZeroKnowledgeClassical2025}}};
    \draw (4.05,1.8) node{{\scriptsize (2025)}};

    \draw (5,2.4) node{{\scriptsize \cite{zhangZeroKnowledgeProofs2020a}}};
    \draw (5.05,2.1) node{{\scriptsize (2020)}};
    \draw (6,3.1) node{{\scriptsize \cite{liu2021zkcnn}}};
    \draw (6.05,2.8) node{{\scriptsize (2021)}};
    \draw (6.5,4) node{{\scriptsize \cite{wangEfficientZeroKnowledgeClassical2025}}};
    \draw (6.55,3.7) node{{\scriptsize (2025)}};
    
    \draw (5.2,3.7) node{{\scriptsize \cite{toreiniFairnessServiceFaaS2024}}};
    \draw (5.25,3.4) node{{\scriptsize (2024)}};
    
    \end{tikzpicture}
    \caption{Emerging structure of ZKML contributions, showing convergence toward a unified framework that supports verification and trustworthy AI.}
    \label{fig:zkml-convergence}
\end{figure}
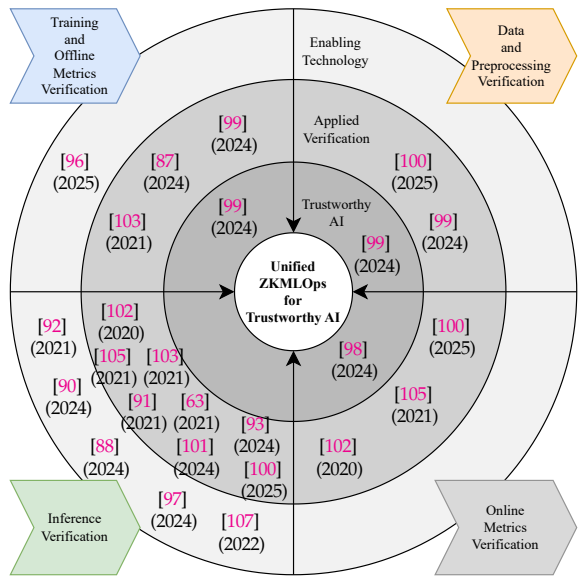

To illustrate the emerging structure of ZKP-Enhanced ML research, we adapted the visualization style of the Thoughtworks Technology Radar\footnote{\url{https://www.thoughtworks.com/insights/blog/build-your-own-technology-radar}}. Figure~\ref{fig:zkml-convergence} highlights how current efforts are concentrated on performance and feasibility, yet indicate a clear trajectory toward trustworthy AI principles. ZKMLOps emerges as the technical foundation for building verifiable, privacy-preserving, and auditable ML systems, thereby enabling the practical realization of trustworthy AI at scale.

\section{Future Work}
\label{sec:future_work}
Future research should prioritize the development of efficient ZKP protocols specifically designed for the data preprocessing and training phases of the machine learning lifecycle. These stages remain critically underexplored compared to the more extensively studied domain of inference verification. Addressing these gaps is essential to enable end-to-end trustworthiness in ML systems.

A valuable avenue for future investigation involves the creation of a decision-support tool, potentially structured as a decision tree, that leverages current state-of-the-art contributions. This tool would assist practitioners in selecting, configuring, and deploying appropriate ZKP techniques tailored to specific use-case requirements, thereby operationalizing ZKMLOps frameworks.

Moreover, comprehensive practical evaluations in real-world settings should be undertaken to assess trade-offs and identify deployment bottlenecks. Empirical studies across diverse application domains can provide insights into the performance, scalability, and regulatory compliance of ZKP-Enhanced ML workflows.

Another promising direction is the analysis of ZKP into federated learning paradigms, where preserving privacy across decentralized and heterogeneous data sources is paramount. Future work should explore how ZKPs can be employed to verify model updates and ensure data integrity without exposing sensitive information or compromising the decentralized architecture of such systems.

By addressing these research priorities, the community can pave the way toward more robust, privacy-preserving, and verifiable AI systems that meet the increasing demands of trust and regulation.

\section{Conclusion}
\label{sec:conclusions}
This study demonstrates the significant potential of ZKPs to enhance verification and validation processes for Trustworthy AI systems, culminating in the conceptualization of a ZKMLOps framework. Our systematic survey and analysis highlight that ZKPs offer cryptographically verifiable and tamper-proof evidence of computational correctness, while preserving the confidentiality of proprietary models and sensitive data.

The identification of five core ZKP properties---interactivity, guarantees, setup requirements, computational representation, and succinctness---provides a robust foundation for their integration into machine learning workflows. Mapping ZKP-enhanced ML applications to the TDSP model reveals a strong research focus on inference verification, underscoring the need for further work in data preprocessing and training phases.

The observed convergence towards a unified ZKMLOps framework reflects an alignment with Trustworthy AI principles such as privacy, accountability, and transparency. This alignment supports compliance with emerging regulations like the EU AI Act and helps cultivate public trust in AI systems, particularly in high-stakes domains.

Future work should address remaining challenges related to protocol efficiency, scalable implementation, real-world evaluation, and integration with federated learning. A decision-support tool tailored to guide practitioners in adopting suitable ZKP methods will further strengthen the operational viability of ZKMLOps pipelines. By advancing these research directions, ZKMLOps can become a standardized, auditable, and privacy-preserving foundation for responsible AI development and deployment in an increasingly regulated and trust-conscious global environment.

\bibliographystyle{IEEEtran}
\bibliography{bibliography} 

\end{document}